\documentstyle[amsfonts,12pt,psfig]{article}
\makeatletter
\def\@cite#1#2{{[{#1}]\if@tempswa\typeout
{IJCGA warning: optional citation argument
ignored: `#2'} \fi}}

\newcount\@tempcntc
\def\@citex[#1]#2{\if@filesw\immediate\write\@auxout{\string\citation{#2}}\fi
  \@tempcnta\z@\@tempcntb\m@ne\def\@citea{}\@cite{\@for\@citeb:=#2\do
    {\@ifundefined
       {b@\@citeb}{\@citeo\@tempcntb\m@ne\@citea\def\@citea{,}{\bf ?}\@warning
       {Citation `\@citeb' on page \thepage \space undefined}}%
    {\setbox\z@\hbox{\global\@tempcntc0\csname b@\@citeb\endcsname\relax}%
     \ifnum\@tempcntc=\z@ \@citeo\@tempcntb\m@ne
       \@citea\def\@citea{,}\hbox{\csname b@\@citeb\endcsname}%
     \else
      \advance\@tempcntb\@ne
      \ifnum\@tempcntb=\@tempcntc
      \else\advance\@tempcntb\m@ne\@citeo
      \@tempcnta\@tempcntc\@tempcntb\@tempcntc\fi\fi}}\@citeo}{#1}}
\def\@citeo{\ifnum\@tempcnta>\@tempcntb\else\@citea\def\@citea{,}%
  \ifnum\@tempcnta=\@tempcntb\the\@tempcnta\else
   {\advance\@tempcnta\@ne\ifnum\@tempcnta=\@tempcntb \else \def\@citea{--}\fi
    \advance\@tempcnta\m@ne\the\@tempcnta\@citea\the\@tempcntb}\fi\fi}
\makeatother
\newenvironment{Eqnarray}%
     {\arraycolsep 0.14em\begin{eqnarray}}{\end{eqnarray}}

\def\be{\begin{equation}}
\def\ee{\end{equation}}
\def\bear{\be\begin{array}}
\def\eear{\end{array}\ee}
\def\bea{\begin{Eqnarray}}
\def\eea{\end{Eqnarray}}

\def\lsim{\mathrel{\raise.3ex\hbox{$<$\kern-.75em\lower1ex\hbox{$\sim$}}}}
\def\gsim{\mathrel{\raise.3ex\hbox{$>$\kern-.75em\lower1ex\hbox{$\sim$}}}}
\def\ifmath#1{\relax\ifmmode #1\else $#1$\fi}
\def\ls#1{\ifmath{_{\lower1.5pt\hbox{$\scriptstyle #1$}}}}

\def\beq{\begin{equation}}
\def\eeq{\end{equation}}
\def\beqa{\begin{Eqnarray}}
\def\eeqa{\end{Eqnarray}}

\def\boxit#1{\leavevmode\thinspace\hbox{\vrule\vtop{\vbox{\hrule%
        \vskip3pt\kern1pt\hbox{\vphantom{\bf/}\thinspace\thinspace%
        {\bf#1}\thinspace\thinspace}}\kern1pt\vskip3pt\hrule}\vrule}%
        \thinspace}
\def\Boxit#1{\noindent\vbox{\hrule\hbox{\vrule\kern3pt\vbox{
        \advance\hsize-7pt\vskip-\parskip\kern3pt\bf#1
        \hbox{\vrule height0pt depth\dp\strutbox width0pt}
        \kern3pt}\kern3pt\vrule}\hrule}}

\def\boxeq#1{\boxit{${\displaystyle #1 }$}}          

\def\baselinestretch{1}
\begin{document}

\catcode`@=11
\newtoks\@stequation
\def\subequations{\refstepcounter{equation}%
\edef\@savedequation{\the\c@equation}%
  \@stequation=\expandafter{\theequation}
  \edef\@savedtheequation{\the\@stequation}
  \edef\oldtheequation{\theequation}%
  \setcounter{equation}{0}%
  \def\theequation{\oldtheequation\alph{equation}}}
\def\endsubequations{\setcounter{equation}{\@savedequation}%
  \@stequation=\expandafter{\@savedtheequation}%
  \edef\theequation{\the\@stequation}\global\@ignoretrue

\noindent}
\catcode`@=12
\setcounter{footnote}{1} \setcounter{page}{1}

\noindent

\title{{\bf Flavour Symmetries and K\"ahler Operators}}
\vskip2in
\author{ 
{\bf J.R. Espinosa$^{1}$\footnote{\baselineskip=16pt E-mail: {\tt
espinosa@makoki.iem.csic.es}. }} and  
{\bf A. Ibarra$^{2}$\footnote{\baselineskip=16pt  E-mail: {\tt
alejandro.ibarra@cern.ch}}} \\ 
\hspace{3cm}\\
 $^{1}$~{\small Instituto de F\'{\i}sica Te\'orica, CSIC/UAM, C-XVI} \\
{\small Universidad Aut\'onoma de Madrid,} \\
{\small Cantoblanco, 28049 Madrid, Spain.}
\hspace{0.3cm}\\
 $^{2}$~{\small Department of Physics, Theory Division, CERN}\\
{\small CH-1211 Geneva 23, Switzerland}
}
\date{}
\maketitle
\def\baselinestretch{1.15}
\begin{abstract}
Any supersymmetric mechanism to solve the flavour puzzle would generate
mixing both in the superpotential Yukawa couplings and in the K\"ahler
potential. In this paper we study, in a model independent way, the impact
of the nontrivial structure of the K\"ahler potential on the physical mixing
matrix, after kinetic terms are canonically normalized. We undertake this
analysis both for the quark sector and the neutrino sector.
For the quark sector, and in view of the experimental values for the
masses and mixing angles, we find that the effects of canonical
normalization are subdominant. On the other hand, for the leptonic sector
we obtain different conclusions depending on the spectrum of neutrinos.  
In the hierarchical case we obtain similar conclusion as in the quark
sector, whereas in the degenerate and inversely hierarchical case, 
important
changes in the mixing angles could be expected.
\noindent

 \end{abstract}

\thispagestyle{empty}

\vskip-17cm
\rightline{IFT-UAM/CSIC-04-21}
\rightline{CERN-PH-TH/2004-082}

\newpage
\baselineskip=20pt

\section{Introduction}

Family replication and flavour dynamics are among the most intriguing
features of Particle Physics \cite{reviews}.
In the Standard Model, three generations of
left-handed quarks, that transform as doublets under the weak interactions,
form Dirac pairs with three generations of right-handed quarks,
that transform as singlets. After the electroweak symmetry breaking,
a $3\times 3$ Dirac mass matrix arises, that is diagonalized by
certain mass eigenstates. Experimentally,  the mass eigenstates
turn out to be mixtures of the weak eigenstates, which mixing
is described by a $3\times 3$ unitary matrix,
the Cabibbo-Kobayashi-Maskawa (CKM) matrix \cite{CKM}.

By observing different weak decay processes
and using experiments of deep inelastic neutrino scattering, 
it has become possible to determine this mixing matrix 
with a fairly high accuracy \cite{PDG}.
The data indicate a hierarchy in the quark mixing angles, that can be
conveniently emphasized using the Wolfenstein parametrization 
\cite{Wolfenstein:1983yz}
\bea
V_{CKM}= \pmatrix{1-{\displaystyle{\lambda^2\over 2}} & \lambda &
A\lambda^3(\rho-i\eta)\cr
&&\cr
-\lambda & 1- {\displaystyle{\lambda^2\over 2}} & A\lambda^2 \cr
&&\cr
A\lambda^3(1-\rho-i\eta) & -A\lambda^2 & 1 \cr} + {\cal O}(\lambda^4) ,
\label{Wolfenstein}
\eea
where $\lambda$ is determined with a very good precision in
semileptonic $K$ decays, giving $\lambda \simeq 0.23$, and $A$ is measured
in semileptonic $B$ decays, giving $A\simeq 0.82$. The parameters
$\rho$ and $\eta$ are more poorly measured, but a rough
estimate is $\rho\simeq 0.1$, $\eta\simeq 0.3$ \cite{Battaglia:2003in}.
In the Wolfenstein parametrization, the hierarchy arises 
from the increasing powers of the different elements in 
the small expansion parameter $\lambda$.

The quark masses are also observed to be hierarchical, and this hierarchy
can be expressed in powers of the same expansion parameter $\lambda$.
Namely:
\bea
m_u:m_c:m_t &\sim& \lambda^8:\lambda^4:1\ , \nonumber \\
m_d:m_s:m_b &\sim& \lambda^4:\lambda^2:1.
\eea

The recent discovery and confirmation of neutrino oscillations
has added a new perspective to the flavour puzzle
\cite{Gonzalez-Garcia:2002dz}. Different
experiments indicate that neutrino mass
eigenstates are also flavour-mixed, resulting in
a leptonic version of the CKM matrix, the 
Maki-Nakagawa-Sakata (MNS) matrix \cite{Maki:mu}
\footnote{With two extra phases when neutrinos are Majorana particles.}.
In stark contrast to the quark case, neutrino mixing angles are not small. 
Neutrino experiments point to a
maximal or nearly maximal atmospheric angle (best fit value,
$\sin^2\theta_{23}=0.52$), a large solar angle (best fit value, 
$\sin^2\theta_{12}=0.30$), and a small 13 angle 
($\sin^2\theta_{13}\lsim 0.053$ @ $3 \sigma$)
\cite{Maltoni:2003da}. On the other hand, 
our present knowledge of the leptonic masses 
has two opposite sides: whereas the
charged-lepton masses are known with an astounding precision,
and they also follow a hierarchical pattern,
\bea
m_e:m_\mu:m_\tau &\sim& \lambda^{4-5}:\lambda^2:1\ , 
\eea
not much is known about neutrino masses. It is known
that the atmospheric mass splitting is $\Delta m^2_{atm}
\simeq 2.6 \times 10^{-3}$ eV$^2$, and the solar mass splitting, 
$\Delta m^2_{sol} \simeq 6.9 \times 10^{-5}$ eV$^2$.
However, it is not known the actual mass spectrum (whether it is
degenerate, hierarchical or inverse hierarchical) or the absolute
scale of neutrino masses. In the hierarchical case one would have
\bea
m_{\nu_1}:m_{\nu_2}:m_{\nu_3} &\sim& \lambda^x:\lambda:1\ ,
\eea
with $x$ arbitrarily high for $m_{\nu_1}\rightarrow 0$. Nothing is known 
either about
CP violation in the leptonic sector.

It is indeed frustrating that despite the large amount of data available, 
a compelling theoretical scenario to explain these data is still lacking.  
In the Standard Model, fermion masses and mixing angles have their origin
in the Yukawa couplings, that are parameters within this model. Therefore,
the origin of flavour has to lie in the realm of physics 
beyond the Standard Model. 

There are though some interesting proposals to explain the origin of these
flavour patterns. Perhaps the most elegant is the Froggatt-Nielsen
mechanism \cite{Froggatt:1978nt}:  left-handed and right-handed quarks of
different generations carry different charges under a flavour symmetry.
This forbids the appearance of some Yukawas which are only generated
through non-renormalizable operators (suppressed by some heavy scale $M$) 
that involve one or more scalar 
fields ($\varphi$), usually called flavons. When these scalar fields 
take a vacuum expectation value the family symmetry is broken 
spontaneously. Assuming that  $\langle\varphi\rangle/M$ is
 of order $\lambda$ then Yukawa couplings naturally small (in the 't Hooft 
sense) are generated, with a non-trivial pattern of masses and 
mixing angles depending on charge assignments.

As explained above, one of the assumptions is that the family symmetry 
breaking occurs at very high energies and involves superheavy fields.  
The presence of these superheavy fields jeopardizes the lightness of the 
Higgs mass, and therefore supersymmetric versions of the Froggatt-Nielsen 
mechanism are more plausible than their non-supersymmetric counterparts.
Many models along these lines exist in the literature, both for abelian
\cite{Leurer:1992wg,Ibanez:ig,Dudas:1995yu,Binetruy:1996xk,Dreiner:2003hw,Dreiner:2003yr}
as for non abelian groups \cite{Barbieri:1997tu,King:2001uz}. The addition 
of supersymmetry constrains even more the flavour pattern of Yukawa 
matrices: some entries that would be allowed in non-supersymmetric 
theories can now be forbidden by the holomorphicity of the superpotential
 (these are the so-called supersymmetric zeros).

A supersymmetric theory is defined by three functions: the 
superpotential, $W(\phi)$, the K\"ahler potential, $K(\phi,\phi^*)$, 
and the gauge kinetic functions, $f_a(\phi)$. Here, $\phi$ represents 
all the chiral matter superfields.
Any supersymmetric mechanism for generating flavour would leave an imprint
both in the superpotential, through the Yukawa couplings, 
and in the K\"ahler potential. In principle, both sources of flavour 
violation contribute to the CKM matrix. 
Taking just the superpotential, ignoring the non-trivial 
structure of the K\"ahler potential, one could 
compute the CKM matrix as the misalignment between the up 
and the down components of the quark doublet, when 
one goes from the weak eigenstate
basis to the mass eigenstate basis. However, the supersymmetric 
theory also comprises the K\"ahler potential and should not be disregarded.
The non-trivial structure of the K\"ahler potential would translate into
non-canonical kinetic terms for quarks, and therefore these fields are non 
physical. The CKM matrix computed in this way is also non physical,
and we will refer to it as the ``naive'' CKM matrix. The correct
procedure requires the consistent redefinition of the quark superfields
to render the kinetic terms canonical \cite{canonical}. This procedure 
will yield the ``physical'' CKM matrix \cite{canonical2}. The purpose of 
this paper is to compare, in a model independent way, the physical CKM 
matrix and the ``naive'' one and study what implications follow.

In Section 2 we derive a very useful and compact formula that relates the
physical CKM matrix to the naive one. In Section 3 we use that formula,
taking full advantage of the hierarchy of quark masses, for a perturbative
analysis of the physical CKM matrix and derive approximate formulas for
the elements of this matrix in terms of the elements of the naive CKM
matrix. In section 4 we illustrate our method with a particular
example and discuss the physical interpretation of our results.
In Section 5 we consider the effects of canonical normalization in
the leptonic sector. The conclusions are summarized in Section 6.

\section{The naive {\it\bf vs.} the physical CKM matrix}

The part of the superpotential relevant for quark mass matrices is
\bea
W_{\rm quark}=Y^u_{ij} Q_i u_j H_2 + Y^d_{ij} Q_i d_j H_1\ ,
\label{superpotential}
\eea
where $Q_i$ ($i=1,2,3$), denote the left-handed quark doublets, $u_i$ 
($d_i$) the up (down)-type right-handed quarks, and $H_2$ ($H_1$) the
up (down)-type Higgs doublets. To fix the notation,
the Yukawa couplings are diagonalized as
\bea 
Y^u &=& V_{u_L} D_u V_{u_R}^{\dagger}\nonumber\\
Y^d &=& V_{d_L} D_d V_{d_R}^{\dagger}\ .
\eea
Disregarding the K\"ahler potential, one could compute the CKM matrix
from the matrices that diagonalize the Yukawa couplings from the left.
As explained in the introduction, this procedure is not complete in 
general, and the result would be the ``naive'' CKM matrix, 
\be
V_{CKM}= V_{u_L}^{\dagger} V_{d_L}\ . 
\ee

To compute the physical CKM matrix, one has to tackle properly the K\"ahler
potential. A general K\"ahler potential reads:
\bea
K=K^Q_{{\bar i}j} Q^{\bar i *} Q^j + K^u_{{\bar i}j} u^{\bar i *} u^j
+ K^d_{{\bar i}j} d^{\bar i *} d^j \ . 
\label{kahler}
\eea
where the matrices $K^\phi$ (with $\phi=Q,u,d$) are dimensionless and  
hermitian  (so that $K$
is real). The minimal 
(or canonical) case corresponds to $K^\phi_{ij}=\delta_{ij}$. In 
general these matrices are functions of other chiral fields, e.g. the 
flavons, which might enter through non-renormalizable operators. 
If such fields take vacuum expectation values, or if 
$K^\phi_{ij}\neq\delta_{ij}$ already at the renormalizable level,  
non canonical kinetic terms would follow. Therefore, a superfield 
redefinition has to be performed in order to get to canonical kinetic 
terms. If $K^Q$ is diagonalized as
\be
K^Q=U_Q^{\dagger} D_K^Q U_Q\ ,
\ee
where $U_Q$ is unitary and $D_K^Q$ diagonal,
and similarly for $K^u$ and $K^d$, then the redefined superfields 
\bea 
\label{redefinitions}
Q' &=& [U_Q^{\dagger} (D_K^Q)^{1/2} U_Q] Q \equiv  V_{Q'Q}Q, \nonumber 
\\
u' &=& [U_u^{\dagger} (D_K^u)^{1/2} U_u] u \equiv  V_{u'u}u, \\
d' &=& [U_d^{\dagger} (D_K^d)^{1/2} U_d] d \equiv  V_{d'd}d\nonumber
\eea
give rise to a canonical K\"ahler potential 
$K={Q'_i}^* {Q'}^i+ {u'_i}^* {u'}^i+{d'_i}^* {d'}^i$. (Notice that 
we have defined the canonically normalized superfields such that 
$\phi'\rightarrow \phi$ as $D_K^{\phi}\rightarrow 1$, with 
$\phi$ any chiral field).

The corresponding superpotential written down in terms of the 
redefined superfields reads\footnote{In our analysis we will ignore
the effects of the canonical normalizations of the Higgs doublets, since
their only effect is a flavour independent scaling of the Yukawa entries
that does not affect the mixing angles.}:
\bea
W_{\rm quark} = 
{Y_{ij}'}^u {Q'_i} {u'_j} H_2+
{Y'_{ij}}^d {Q'_i} {d'_j} H_1\,
\label{newW}
\eea
where
\bea
{Y'}^u&=&V_{QQ'}^T Y^u V_{uu'}, \nonumber \\
{Y'}^d&=&V_{QQ'}^T Y^d V_{dd'}\ ,
\label{newYuks}
\eea
are the physical Yukawa couplings and 
\be
V_{\phi\phi'}\equiv V_{\phi'\phi}^{-1}=U_\phi^{\dagger} (D_K^\phi)^{-1/2} 
U_\phi\ .
\ee 
 These physical Yukawa couplings
can be diagonalized from the left by ${V'_{u_L}}$ and 
${V'_{d_L}}$, and the misalignment between these two matrices
yields the physical CKM matrix: ${V'_{CKM}}= 
{{V'_{u_L}}^{\dagger}} {V'_{d_L}}$. 

It is not a trivial task to relate, in a model independent way,
the ``naive'' CKM matrix to the physical one. To this end, 
we first rewrite 
eq.~(\ref{newYuks}) as ${Y'}^u=Z_{u_L} D_u Z_{u_R}^{\dagger}$,
and similarly for ${Y'}^d$.
This looks like a singular value decomposition, but the matrices
\bea
Z_{u_L}&=&V_{QQ'}^T V_{u_L}= [U_Q^{T} (D_K^Q)^{-1/2} U_Q^*] V_{u_L},  
\nonumber \\
Z_{u_R}&=&V_{uu'}V_{u_R} = [U_u^{\dagger} (D_K^u)^{-1/2} U_u] V_{u_R}
\label{matrixZ}
\eea
are not unitary in general for $D_K^Q\neq I$. On the other hand, it is 
straightforward to 
construct unitary matrices from $Z_{u_L}$ and $Z_{u_R}$,
using the Gram-Schmidt procedure. Given the column vectors of the 
matrix $Z = (z_1,z_2,z_3)$, the procedure
guarantees that the following set of column vectors is orthonormal:
\bea
\label{GS}
w_3 &=& \frac{z_3}{N_3}, \nonumber \\ 
w_2 &=& \frac{z_2-\langle w_3,z_2 \rangle w_3}{N_2},  \\
w_1 &=& \frac{z_1-\langle w_2,z_1 \rangle w_2-
\langle w_3,z_1 \rangle w_3}{N_1},  \nonumber
\eea
where $N_1$, $N_2$ and $N_3$ are normalization factors, and 
$\langle ~,~ \rangle$ denotes the scalar product in ${\mathbb C}^3$.
(Subindices $u_{L,R}$ or $d_{L,R}$ should be understood everywhere in these
formulas.) In matricial form, eqs.~(\ref{GS}) can be cast as $Z=W T$, with
$W$ a unitary matrix with column vectors $W=(w_1,w_2,w_3)$ 
and $T$ a lower triangular matrix:
\bea
T=\pmatrix{ N_1 & 0 & 0 \cr
 \langle w_2,z_1 \rangle & N_2 & 0 \cr
\langle w_3,z_1 \rangle & \langle w_3,z_2 \rangle & N_3} .
\eea
Substituting these matricial forms of $Z_{u_L,d_L}$ in 
${Y'}^u$ we get ${Y'}^u=W_{u_L}T_{u_L}D_u T_{u_R}^{\dagger} 
W_{u_R}^{\dagger}$. Finally, we diagonalize 
\be
T_{u_L}D_u T_{u_R}^{\dagger}=
R_{u_L} {D'_u} R_{u_R}^{\dagger}\ ,
\label{intermediate}
\ee
with $R_{u_L}$ and 
$R_{u_R}$ unitary, to obtain 
\bea
{Y'}^u=W_{u_L} R_{u_L} {D'_u} 
R_{u_R}^{\dagger} W_{u_R}^{\dagger} ,
\eea
which is nothing but the singular value decomposition of the physical
Yukawa coupling ${Y'}^u={V'_{u_L}}
{D'_u}{V'_{u_R}}^{\dagger}$, with 
${V'_{u_{L,R}}}=W_{u_{L,R}} R_{u_{L,R}}$. Note that
${V'_{u_{L,R}}}$ are guaranteed to be unitary by 
construction. This diagonalization
procedure in two steps has the advantage over the direct diagonalization
${Y'}^u={V'_{u_L}} {D'_u}{V'_{u_R}}^{\dagger}$, that it will enable us to 
write down the physical CKM matrix in terms of the ``naive'' one in a simple 
way. Also, this
factorization will prove to be useful later on, since the off-diagonal
entries of $R_u$ turn out to be suppressed by ratios of quark
masses.

For quark masses, we can write a relation between the physical and naive 
Yukawa eigenvalues using the (non-singular, non-unitary) matrices
\be
S_\phi\equiv R_\phi^\dagger T_\phi\ ,
\label{Smatrix}
\ee
(where, as usual, $\phi=u_L,u_R,d_L,d_R$) in eq.~(\ref{intermediate}).  
This relation is simply
\be
\boxeq{ D'_u=S_{u_L} D_u S_{u_R}^\dagger\ }
\label{newmasses}
\ee
and a similar one for the down quarks. From these equations it follows 
that the physical eigenvalues are, keeping only the dominant 
terms,
\bea
{m'_u} &\simeq& N^{u_L}_1 N^{u_R}_1 m_u \nonumber \\
{m'_c} &\simeq& N^{u_L}_2 N^{u_R}_2 m_c  \\
{m'_t} &\simeq& N^{u_L}_3 N^{u_R}_3 m_t \nonumber
\eea
and similarly for the down-quark sector. Canonization of the K\"ahler
potential just changes the masses by a normalization factor, that
in many models is very close to one. A related conclusion from 
(\ref{newmasses}) is that if some quark is massless before taking into 
account K\"ahler corrections, it will remain so afterwards.

On the other hand, the physical CKM matrix is:
\bea
{V'_{CKM}}={V'}_{u_L}^{\dagger}{V'_{d_L}}=
R_{u_L}^{\dagger} W_{u_L}^{\dagger} W_{d_L} R_{d_L} .
\eea
The ``naive'' CKM matrix enters implicitly in the right-hand 
side of this equation. To make it explicit, we use  
$W_{u_L}^{\dagger}=W_{u_L}^{-1}$, $W=Z T^{-1}$ 
and eq.~(\ref{matrixZ}) to obtain
\bea
{V'_{CKM}}=R_{u_L}^{\dagger} T_{u_L} V_{CKM} T_{d_L}^{-1} R_{d_L}.
\label{newCKM}
\eea
Making use again of the matrices $S_\phi$ defined in 
eq.~({\ref{Smatrix}}) we find the central formula of the paper
\be
\boxeq{
V'_{CKM}=S_{u_L} V_{CKM} S_{d_L}^{-1} }
\label{newformCKM}
\ee
Eqs.~(\ref{newmasses},\ref{newformCKM}) show explicitly that the effect of a 
non-canonical K\"ahler potential on quark masses and mixing angles is
that of an equivalence transformation, as given by the matrices $S_\phi$.

\section{Perturbative analysis of the physical CKM matrix}

All the formulas that we have presented so far are exact. To proceed
we have to compute analytically the matrices $R_{u_L,d_L}$
introduced in eq.~(\ref{intermediate}).
This cannot be done exactly, but as we will show explicitly later on, 
$(R_{d_L})_{12}\lsim m_d/m_s \sim \lambda^2$, $(R_{d_L})_{13}\lsim 
m_d/m_b\sim \lambda^4$, 
$(R_{d_L})_{23}\lsim m_s/m_b\sim \lambda^2$, and similarly for the up-type 
rotations: $(R_{u_L})_{12}\lsim m_u/m_c \sim \lambda^4$, 
$(R_{u_L})_{13}\lsim
m_u/m_t\sim \lambda^8$,
$(R_{u_L})_{23}\lsim m_c/m_t\sim \lambda^4$. 
Therefore, the smallness of the ratios of the quark masses
allows a perturbative expansion of $R_{u_L,d_L}$ and of the physical
CKM matrix. At this point it is convenient to point out that
the Gram-Schmidt procedure used in the previous section is not unique. 
The particular ordering of vectors used in eq.~(\ref{GS}) is important
for the success of the perturbative analysis carried out in this 
section because it starts by treating $z_3$ as a good approximation to the 
eigenvector with the heaviest eigenvalue (which is reasonable because 
corrections will be suppressed by the smallness of the other two 
eigenvalues), then proceeds with the next-to-heaviest eigenvector and so 
on.

We start computing the zero-th order approximation to $V'_{CKM}$, i.e. we
neglect any effect coming from the matrices $R_{u_L,d_L}$ which we 
approximate by the identity. The physical CKM matrix reads, at zero-th 
order
\bea
V'^0_{CKM}=T_{u_L} V_{CKM} T_{d_L}^{-1}.
\label{CKM-0}
\eea
Due to the triangular structure of the $T$ matrices, the elements
above the main diagonal of ${V'}^0_{CKM}$ 
have a fairly simple expression:
\bea
\label{zero-order}
(V'^0_{CKM})_{12}&=&\frac{N^{u_L}_1}{N^{d_L}_2} {(V_{CKM})}_{12} 
-\frac{N^{u_L}_1}{N^{d_L}_2} 
\frac{\langle w^{d_L}_3,z^{d_L}_2 \rangle}{N^{d_L}_3} {(V_{CKM})}_{13},  
\nonumber \\
(V'^0_{CKM})_{13}&=&\frac{N^{u_L}_1}{N^{d_L}_3} 
{(V_{CKM})}_{13}, \\
(V'^0_{CKM})_{23}&=&\frac{N^{u_L}_2}{N^{d_L}_3} {(V_{CKM})}_{23} 
+ \frac{\langle w^{u_L}_2,z^{u_L}_1 \rangle}{N^{d_L}_3} {(V_{CKM})}_{13} .
\nonumber
\eea
It is possible to compute explicitly the normalization factors,
as well as the scalar products $\langle w,z \rangle$ in terms
of the K\"ahler potential and the matrices that diagonalize the 
Yukawa couplings in the original superpotential, $V_{u_L,d_L}$.
Defining $Q_{\phi}\equiv V_{\phi}^{\dagger} {(K^{\phi})}^{-1} V_{\phi}$ and 
noting that $\langle z^{\phi}_i,z^{\phi}_j \rangle=(Q_{\phi})_{ij}$ we 
obtain
\bea
\label{Qs}
\langle w^{\phi}_2,z^{\phi}_1 \rangle = 
-\frac{1}{N^{\phi}_2}\frac{(Q^{-1}_{\phi})_{12}}{(Q_{\phi})_{33}}
{\rm det} Q^{\phi}, ~~~&&~~~
(N^{\phi}_1)^2=\frac{1}{(Q^{-1}_{\phi})_{11}}, \nonumber \\
\langle w^{\phi}_3,z^{\phi}_1 \rangle = 
\frac{(Q_{\phi})_{31}}{N^{\phi}_3},  
~~~&&~~~
(N^{\phi}_2)^2=\frac{(Q^{-1}_{\phi})_{11}}{(Q_{\phi})_{33}}~{\rm det} 
Q^{\phi}, \\
\langle w^{\phi}_3,z^{\phi}_2 \rangle = \frac{(Q_{\phi})_{32}}
{N^{\phi}_3}, ~~~&&~~~
(N^{\phi}_3)^2=(Q_{\phi})_{33}, \nonumber
\eea
where $\phi$ represents $u_{L,R}$ and $d_{L,R}$.

Motivated by flavour symmetries, we find reasonable 
to assume that the K\"ahler  is a perturbation from
the identity. Hence, the diagonal elements of $Q_\phi$
are very close to one, and the off-diagonal elements
are suppressed with respect to those in the diagonal. If
this is the case, all the normalization factors are 
approximately one and all the scalar products are much smaller 
than one\footnote{Even in the case in which there are large
mixings in $V_{u_L}$, as in democratic models \cite{Harari:1978yi}, 
the off-diagonal elements of the matrix $Q_\phi$ are of the order 
of the perturbation itself.}.

From eq.~(\ref{zero-order}) one realizes that, at zero-th order, the 13 
angle
in the CKM matrix does not change substantially, at most by a factor that is 
close to one. The equation for $({V'}^0_{CKM})_{13}$
also tells us that, barring cancellations, the 13 element of the ``naive''
CKM matrix, $({V_{CKM}})_{13}$ should not be larger than $\sim \lambda^3 
A$.

Concerning the 12  and 23 angles, since the observed 
values are respectively of order $\lambda$ and $\lambda^2 A$, 
the ${(V_{CKM})}_{13}$ contribution in eq.~(\ref{zero-order}),
being at most of order $\sim \lambda^3 A$, is irrelevant. 
Therefore, to zero order in the perturbation
expansion in ratios of quark masses, none of the CKM elements changes
substantially:
\bea
(V'^0_{CKM})_{12}&\simeq&
\frac{N^{u_L}_1}{N^{d_L}_2} {(V_{CKM})}_{12}, \nonumber \\
(V'^0_{CKM})_{13}&\simeq&
\frac{N^{u_L}_1}{N^{d_L}_3} {(V_{CKM})}_{13}, \\
(V'^0_{CKM})_{23}&\simeq&
\frac{N^{u_L}_2}{N^{d_L}_3} {(V_{CKM})}_{23}. \nonumber
\eea

Let us analyze now the first-order terms in the expansion.
To this order, the  off-diagonal
elements in the rotation matrices $R_{u_L}$ and $R_{d_L}$ read:
\bea
\label{rots}
(R_{d_L})_{12}\simeq \frac{m_d}{m_s} \frac{N^{d_L}_1}{N^{d_L}_2 N^{d_R}_2}
\langle  z^{d_R}_1,w^{d_R}_2 \rangle,  \nonumber \\
(R_{d_L})_{13}\simeq \frac{m_d}{m_b} \frac{N^{d_L}_1}{N^{d_L}_3 N^{d_R}_3}
\langle  z^{d_R}_1, w^{d_R}_3 \rangle, \\
(R_{d_L})_{23}\simeq \frac{m_s}{m_b} \frac{N^{d_L}_2}{N^{d_L}_3 N^{d_R}_3}
\langle  z^{d_R}_2, w^{d_R}_3 \rangle, \nonumber 
\eea
and similarly for the up-quark sector.
The explicit expressions for the scalar products
and the normalization factors can be read from eq.~(\ref{Qs}).
An order of magnitude estimate
of these matrix elements is, in view of the observed hierarchy of masses,
\bea
\label{rots-numbers}
(R_{d_L})_{12}\lsim \lambda^2, ~~~&&~~~(R_{u_L})_{12}\lsim \lambda^4, 
\nonumber\\
(R_{d_L})_{13}\lsim \lambda^4, ~~~&&~~~(R_{u_L})_{13}\lsim \lambda^8,\\
(R_{d_L})_{23}\lsim \lambda^2, ~~~&&~~~(R_{u_L})_{23}\lsim \lambda^4,\nonumber
\eea
since the $\langle w,z \rangle$'s in these equations depend on some 
off-diagonal terms
of the K\"ahler potential, and are supposed to be smaller than one.
Being the observed CKM angles of the order of $\lambda$, $\lambda^2 A$ and
$\lambda^3 A (\rho-i\eta)$, it is apparent 
that the only rotations that could have some
impact on the CKM angles are $(R_{d_L})_{12}$ and $(R_{d_L})_{23}$;
the rest are smaller than $\sim \lambda^4$. 
Furthermore, all the remaining terms 
in the series expansion are suppressed at least  by $\lambda^4$, and can 
be neglected.

To quantify the effect of these contributions we use 
eqs.~(\ref{newCKM},\ref{CKM-0}) to write
\bea
{V'_{CKM}}=R_{u_L}^{\dagger} V'^0_{CKM} R_{d_L}.
\label{CKM-2}
\eea
The 12 element in ${V'_{CKM}}$ is of order $\lambda$, so
the rotations in eq.~(\ref{rots-numbers})  have only
a subdominant effect. At the end of the day, the large size
of the Cabibbo angle compared to the hierarchy of quark masses,
implies that the dominant contribution to the quark flavour violation in 
the 12 sector must come from the Yukawa couplings, being the contributions
from the K\"ahler potential subdominant. In other words, 
if one wants to compute the 12 angle, it is not necessary to 
normalize canonically the quark superfields: 
computing the Cabibbo angle from the initial Yukawa couplings, 
eq.~(\ref{superpotential}), is going to give essentially the correct 
result:
\bea
{({V'_{CKM}})}_{12}\simeq({V_{CKM}})_{12} .
\label{physical12}
\eea

The case of the 13 angle, that experimentally is 
${({V'_{CKM}})}_{13}\simeq \lambda^3 A 
(\rho-i \eta)$ requires a more careful analysis. Following 
eqs.~(\ref{CKM-2}) and (\ref{rots}), we obtain:
\bea
{({V'_{CKM}})}_{13}\simeq(V'^0_{CKM})_{13}
+\frac{m_s}{m_b}[(V_{d_R}^{\dagger} K^d V_{d_R})^{-1}]_{23} 
(V'^0_{CKM})_{12}.
\label{physical13}
\eea
Note that rotations in the right-handed sector {\it do} contribute to
the CKM matrix, although this contribution is suppressed
by small ratios of quark masses. The second term in that formula
is $\sim \lambda^3 
[(V_{d_R}^{\dagger} K^d V_{d_R})^{-1}]_{23}$ and 
only in the case in which the off-diagonal element
of the K\"ahler potential is not very suppressed, this term could 
contribute substantially to the physical 13 angle (see below). In general,
$K^d$ departs from the identity through suppressed 
contributions of non-renormalizable origin and therefore there is an 
additional suppression through the off-diagonal element  $
[(V_{d_R}^{\dagger} K^d V_{d_R})^{-1}]_{23}$.

Finally, for the 23 angle, which experimentally is 
${(V'_{CKM})}_{23} \sim \lambda^2 A$,
we obtain:
\bea
{(V'_{CKM})}_{23}\simeq(V'^0_{CKM})_{23}+
\frac{m_s}{m_b} [(V_{d_R}^{\dagger} K^d V_{d_R})^{-1}]_{23}.
\label{physical23}
\eea
The second term is $\sim \lambda^2 
[(V_{d_R}^{\dagger} K^d V_{d_R})^{-1}]_{23}$,
and it could have some effect on the observed 23 angle, depending
again on the K\"ahler potential for the down right-handed quarks.
Nevertheless, if the off-diagonal terms of the K\"ahler potential are
smaller than $\sim \lambda$, that term is at most
$\sim \lambda^3$ and can also be neglected. 

Let us consider also the case in which the K\"ahler potential is not
a perturbation from the identity. As mentioned in the last
paragraph, in this case the K\"ahler potential could contribute
to the 23 angle. However, if this were the case, that same element
of the K\"ahler potential would also contribute to the 13 angle, 
giving ${(V'_{CKM})}_{13}\simeq {({V'_{CKM}})}_{23}
{({V'_{CKM}})}_{12} \simeq \lambda^3 A$, which is too large
by a factor of three. Hence, the 23 angle has to come from the Yukawa
couplings, otherwise the 13 angle would be too large. Note that the 
13 angle cannot be made smaller by cancellations, since all the 
remaining terms are at least of order $\lambda^4$,
or by radiative corrections, since these affect also the 13 angle
in such a way that the relation ${({V'_{CKM}})}_{13}\simeq 
{({V'_{CKM}})}_{23} {({V'_{CKM}})}_{12} $ 
is approximately scale independent \cite{Olechowski:1990bh}.

From the above discussion, it is apparent that the 12 and 23 angles cannot
have their origin in the K\"ahler potential. The 13 angle could,
if $[(V_{d_R}^{\dagger} K^d V_{d_R})^{-1}]_{23}$ is not very suppressed.
This would suggest some approximate symmetry between the second
and third generations of right-handed down quarks. This symmetry
could also generate Yukawa entries large enough to produce
a 13 angle in the ``naive'' CKM matrix of the same order
of magnitude, or larger, than the contribution from the K\"ahler potential.
For instance, in the context of a $U(1)$ symmetry, if the
charges for ${d_R}_2$ and  ${d_R}_3$ are identical, a large 23 entry
in the K\"ahler
potential is allowed by the symmetry. However, it is easy to check
that if one wants to reproduce the Cabibbo angle and the
correct ratios of quark masses, these charges imply a 13 element
in the ``naive'' CKM matrix of $\sim \lambda^3$, and therefore as large
as the contribution from the K\"ahler potential itself. In this very
special case, the effects of the K\"ahler potential are also subdominant.

\section{Froggatt-Nielsen symmetry}

As we have mentioned already, supersymmetric Froggatt-Nielsen models can
easily explain zeros in the Yukawa matrices by combining the flavour
symmetry with the holomorphicity of the superpotential $W$. In this
context, corrections to the Yukawa textures from the K\"ahler potential
could play a very relevant role: couplings that were forbidden in $W$
could be written in $K$ with the final effect of lifting at some order the
supersymmetric zeros of the Yukawa textures. Family symmetries that would
have been otherwise excluded might be viable after all
\cite{Dudas:1995yu}.

In fact, it is true that K\"ahler corrections can change significantly the
Yukawa textures and lift some of the supersymmetric zeros in them.
However, as we have shown in the previous section, the dominant effect
comes from $K^Q$ which is the same [by $SU(2)_L$ gauge invariance] for
$u_L$ and $d_L$. This means that the change induced by the K\"ahler
corrections affect the textures for $Y^u$ and $Y^d$ in a very correlated
way and the net effect on $V_{CKM}$ cancels out. The effects coming from
$K^u$ and $K^d$, the K\"ahler for the right-handed quarks, have to pay the
price of a quark mass in order to propagate to the left-handed sector and
this makes them subdominant.

We give now a concrete example, taken from \cite{Dudas:1995yu}, to 
illustrate this behaviour. It is a model with a $U(1)_F$ flavour symmetry
and a single flavon field $\varphi$, with $F$-charge $-1$. The $F$-charges
of the quarks are such that $Q_{13}\equiv Q_1-Q_3=-2$, $Q_{23}=-3$,
$u_{13}=10$, $u_{23}=7$, $d_{13}=6$ and $d_{23}=5$. With these charges
the Yukawa textures are
\bea
Y^u\sim \left[\begin{array}{ccc}
\lambda^8 & \lambda^5 & 0\\
\lambda^7 & \lambda^4 & 0\\
\lambda^{10} & \lambda^7 & 1
\end{array}\right]\ ,\;\;\;
Y^d\sim \left[\begin{array}{ccc}
\lambda^4 & \lambda^3 & 0\\
\lambda^3 & \lambda^2 & 0\\
\lambda^6 & \lambda^5 & 1
\end{array}\right]\ ,
\eea
with two supersymmetric zeros each. The naive CKM that follows is
\bea
V_{CKM}\simeq \left[\begin{array}{ccccc}
1-\lambda^2/2& & \lambda & &\lambda^8\\
-\lambda & &1-\lambda^2/2 & & \lambda^7\\
-\lambda^8 & & -\lambda^7 & & 1
\end{array}\right]\ ,
\eea
with too small values of $V_{ub}$ and $V_{cb}$. 

For the K\"ahler matrices we have the textures
\bea
K^Q\sim \left[\begin{array}{ccc}
1 & \lambda & \lambda^2\\
\lambda & 1 & \lambda^3\\
\lambda^2 & \lambda^3 & 1
\end{array}\right]\ ,\;\;\;
K^d\sim \left[\begin{array}{ccc}
1 & \lambda & \lambda^6\\
\lambda & 1 & \lambda^5\\
\lambda^6 & \lambda^5 & 1
\end{array}\right]\ ,\;\;\;
K^u\sim \left[\begin{array}{ccc}
1 & \lambda^3 & \lambda^{10}\\
\lambda^3 & 1 & \lambda^7\\
\lambda^{10} & \lambda^7 & 1
\end{array}\right]\ .
\eea
Only $K^Q$, $K^d_{12}$ and $K^u_{12}$ are in principle large enough to be
able to cure the smallness of $V_{ub}$ and $V_{cb}$. In fact $K^Q$ is able
to fill in the zeros in $Y^u$ and $Y^d$ above and gives $Y'^{u,d}_{13}\sim
\lambda^2$ and $Y'^{u,d}_{23}\sim\lambda^3$. The resulting textures, if
uncorrelated, would have produced a CKM matrix much closer to the observed
one, with $V'_{ub}\sim \lambda^2$ and $V'_{cb}\sim\lambda^3$. However, the
correlation makes these matrix elements much smaller. Applying the general
results of the previous section, eqs.~(\ref{physical13},\ref{physical23}),
we find that the K\"ahler contributions to $V'_{ub}$ and $V'_{cb}$ are of
the same order of the naive ones, and thus too small:
\bea
V'_{ub}&\simeq & V_{ub} +\frac{m_s}{m_b}
[(V_{d_R}^{\dagger} K^d V_{d_R})^{-1}]_{23} V_{uc}\sim \lambda^8 
+ \lambda^2 
[\lambda^5] \lambda\sim \lambda^8\ ,\\
V'_{cb}&\simeq & V_{cb} +\frac{m_s}{m_b}
[(V_{d_R}^{\dagger} K^d V_{d_R})^{-1}]_{23} \sim \lambda^7 
+ \lambda^2 [\lambda^5]\sim \lambda^7\ .
\eea
Therefore, this
$U(1)_F$ symmetry has to be rejected (see also \cite{Dreiner:2003yr}).

In the previous example it might seem a coincidence that
$[(V_{d_R}^{\dagger} K^d V_{d_R})^{-1}]_{23}$ happens to have just the
size required to get a contribution of the same order as that of the naive
CKM matrix. However there is a good reason for that. The simplest way of
understanding it is that, by making holomorphic redefinitions of fields,
there are operators in the K\"ahler potential that can be moved to the
superpotential and viceversa. In fact, by such redefinitions one can make
zero all the derivatives $\partial^n K/(\partial \phi^*_{i_1} \partial
\phi_{i_2} \ldots \partial \phi_{i_n})$ ($n>2$) and their conjugates
around a given point \cite{grk} (these are the so-called K\"ahler {\em
normal coordinates}). 
For instance, in the example above, the 23 down-quark sector of the 
K\"ahler potential is of the form
\be
K= |d_2|^2 + |d_3|^2 + {1\over M^5}[d_2 d_3^* \varphi^5 + {\mathrm 
h.c.}]\ .
\ee
The holomorphic redefinition $d_2\rightarrow d_2 - d_3\varphi^5/M^5$
transforms $K$ into
\be
K= |d_2|^2 + |d_3|^2(1 - |\varphi|^{10}/M^{10}) \ .
\ee
Such field redefinitions respect the flavour symmetry and therefore
they do not change the texture of the Yukawas in the superpotential
(which is supposed to be generic and to contain all couplings allowed by 
the flavour symmetry). To summarize, the $K^d_{23}$ element cannot give 
contributions different from those already present through the Yukawa 
couplings because a holomorphic redefinition of fields can be used to 
move such operator from the K\"ahler to the superpotential\footnote{If 
there are at least two flavon fields with charges of opposite signs, such 
holomorphic redefinitions to remove $K^d_{23}$ might not be possible, but 
in such case, $K^d_{23}$ itself will be more suppressed in general.}, 
supposed to be generic to start with.

\section{The lepton sector}

The analysis of the previous sections can be straight-forwardly extended
to the lepton sector. We first discuss the case in which neutrino masses
come from a non-renormalizable dimension five operator 
in the superpotential:
\bea
W_{lep}=Y^e_{ij} L_i e_j H_1+\frac{1}{2}\kappa_{ij} 
(L_i H_2)^T (L_j H_2)
\label{Wlep}
\eea
where $L_i$ denote the left-handed lepton doublets and $e_i$
the right-handed lepton singlets. The matrix $\kappa$, that has
dimensions of mass$^{-1}$, produces neutrino Majorana masses
in the Lagrangian $(1/2) {\cal M}_{\nu} \nu^T \nu$ after
electroweak symmetry breaking, where 
${\cal M}_{\nu}= \kappa \langle H^0_2 \rangle^2$.
The notation for this section is parallel to the one we used in 
Section 2: the charged-lepton Yukawa coupling is diagonalized by 
$Y^e=V_{e_L}D_e V^{\dagger}_{e_R}$ and the dimension five operator
by $\kappa=V_{\nu_L} D_{\kappa} V_{\nu_L}^T$. One could
disregard the effects of the K\"ahler potential and compute 
the MNS matrix from the misalignment
between the charged-lepton and neutrino components of the lepton doublet.
This procedure would yield the ``naive'' MNS matrix,
$V_{MNS}=V^{\dagger}_{e_L} V_{\nu_L}$.

The physical MNS matrix is computed by redefining properly the 
chiral superfields to bring the kinetic terms to their canonical
form. Following the same steps as in Section 2, 
eqs.~(\ref{redefinitions})-(\ref{newformCKM}), one obtains 
the following expresion for the physical
MNS matrix in terms of the ``naive'' one:
\bea
{V'_{MNS}}=S_{e_L} V_{MNS} S_{\nu_L}^{-1}\ ,
\label{newMNS}
\eea
with $S_{e_L}\equiv R_{e_L}^{\dagger} T_{e_L}$ and a similar expression 
for $S_{\nu_L}$.

We first compute the zero-th order in the expansion, {\it i.e.} we
neglect any contribution coming from $R_{e_L}$ and $R_{\nu_L}$:
\bea
\label{zero-order-MNS}
(V'^0_{MNS})_{12}&=&\frac{N^{e_L}_1}{N^{\nu_L}_2} {(V_{MNS})}_{12} 
-\frac{N^{e_L}_1}{N^{\nu_L}_2} 
\frac{\langle w^{\nu_L}_3,z^{\nu_L}_2 \rangle}{N^{\nu_L}_3} {(V_{MNS})}_{13},  
\nonumber \\
(V'^0_{MNS})_{13}&=&\frac{N^{e_L}_1}{N^{\nu_L}_3} 
{(V_{MNS})}_{13}, \\
(V'^0_{MNS})_{23}&=&\frac{N^{e_L}_2}{N^{\nu_L}_3} {(V_{MNS})}_{23} 
+ \frac{\langle w^{e_L}_2,z^{e_L}_1 \rangle}{N^{\nu_L}_3} {(V_{MNS})}_{13} .
\nonumber
\eea

From these equations it is apparent that at order zero the
13 angle does not change substantially, and that, barring cancellations,
the 13 element in the ``naive'' MNS matrix has to be of the same
order of magnitude or smaller than the observed one (we recall
that the 13 angle is bound to be less than $\sim 0.23$ 
at the $3\sigma$ level). On the other hand, the solar and 
atmospheric angles receive contributions from the 13 angle that
are proportional to off-diagonal elements in the K\"ahler potential 
for the lepton doublets. It is interesting to note 
that if the solar angle is maximal in the ``naive'' 
MNS matrix, due for instance to some symmetry
in the parameters of the superpotential, 
the contribution to the mixing from the
K\"ahler potential could explain the observed deviation from maximality, 
provided the 13 angle is close to the experimental upper bound and
$K^L$ has a large off-diagonal entry in the 23 sector. Notice also
that to guarantee that the observed atmospheric angle is close 
to the maximal value, as suggested by experiments, the off-diagonal
entry in the K\"ahler potential in the 12 sector 
would have to be small (again barring cancellations).

The complete analysis of eq.~(\ref{newMNS}), including the effects
of the matrices $R_{e_L}$ and $R_{\nu_L}$ is more involved. This 
requires the diagonalization of $3\times3$ matrices and, although
can be done exactly, the resulting formulas are not very
elucidating. In the quark case, we used the fact that quark masses are very
hierarchical to simplify the formulas. However, the neutrino
spectrum is not known yet. Three possibilities are allowed
experimentally, degenerate, inverted hierarchy and normal hierarchy,
and the analysis should be done separately for each case. We expect 
the analysis for the latter case to be similar to the quark case,
and so would be the conclusions: the effects from the K\"ahler potential
on the mixing angles would be at most of the same order of magnitude
of the ``naive'' mixing angles themselves. On the other hand, 
we expect different conclusions for the 
case in which neutrinos are degenerate or inversely hierarchical.
In this case, $R_{\nu_L}$ could have 
rather large off-diagonal entries that could
contribute significantly to the mixing angles (depending of course
on the structure of the K\"ahler potential). As a matter of fact,
the sensitivity of neutrino mixing angles to extra effects when 
there are degeneracies in the neutrino mass spectrum is a very
well known fact. For instance, when neutrinos are degenerate in mass,
extra effects from radiative corrections can drive the mixing angles to
values much different to the ones that one would naively deduce from the
bare superpotential \cite{RGEnu}.

Here, we will discuss with some detail first the case in which neutrinos 
are hierarchical and then the degenerate case. In the hierarchical case, 
atmospheric and solar neutrino experiments
indicate that the mass of the heaviest neutrino is approximately five
times larger than the mass of the next-to-heaviest neutrino. This is a
rather mild hierarchy, much milder than in the quark sector, 
and could have some impact on the mixing angles.
On the other hand, the mass of the lightest neutrino is unknown --  
experiments are even compatible with a massless lightest neutrino.
For simplicity, we will assume that the mass hierarchy between the
lightest neutrino and the others is very large.

Under these assumptions, only $(R_{\nu_L})_{23,32}$ can give a sizeable
contribution to the leptonic mixing angles; the remaining contributions
are negligible due to the hierarchy in the charged-lepton 
sector and the assumed hierarchy between the lightest
neutrino and the other two neutrinos. Consequently, to first order in 
perturbation theory, the physical mixing angles read
\bea
\label{physicalMNS}
{({V'_{MNS}})}_{12}&\simeq& {(V'^0_{MNS})}_{12}\\
{({V'_{MNS}})}_{13}&\simeq&(V'^0_{MNS})_{13}
+\frac{m_{\nu_2}}{m_{\nu_3}}[(V_{\nu_L}^{\dagger} K^{\nu} 
V_{\nu_L})^{-1}]_{23} 
(V'^0_{MNS})_{12} \\
{({V'_{MNS}})}_{23}&\simeq&(V'^0_{MNS})_{23}+
\frac{m_{\nu_2}}{m_{\nu_3}} [(V_{\nu_L}^{\dagger} K^{\nu} 
V_{\nu_L})^{-1}]_{23}.
\eea

The solar angle at first order does not differ substantially
from the value at zero-th order. How the zero-th order solar angle
is affected by the K\"ahler potential was discussed 
after eq.~(\ref{zero-order-MNS}). On the other hand,
the atmospheric angle is known to be very close to maximal:
$\sin^2\theta_{23}$ lies in the interval 0.31-0.72 at the
$3\sigma$ level, corresponding the best fit point to $0.52$. This suggests
the existence of some underlying symmetry in the superpotential that
yields maximal atmospheric mixing in the ``naive'' MNS matrix. Under 
this assumption, the deviation of the atmospheric angle from
the maximal value is controlled by $(m_{\nu_2}/m_{\nu_3}) 
[(V_{\nu_L}^{\dagger} K^L V_{\nu_L})^{-1}]_{23}$. This combination
also affects the physical 13 angle, and therefore, the contribution
from the K\"ahler potential to the 13 angle is proportional to the
deviation from maximality of the atmospheric angle:
\bea
{({V'_{MNS}})}_{13}&\simeq&
(V'^0_{MNS})_{13}+
[{({V'_{MNS}})}_{23}-(V'^0_{MNS})_{23}]
{({V'_{MNS}})}_{12}\ .
\eea
From this formula one finds an interesting lower bound on the 13 angle:
\bea
\sin^2 \theta_{13}\gsim {1\over 2} 
\sin^2(\theta_{atm}-\pi/4)\sin^2\theta_{sol}\ .
\eea
(We remind that this relation is valid under the assumption 
that the superpotential produces maximal atmospheric mixing by itself.)

In cases with some neutrino mass degeneracy (inverse hierarchy or 
degenerate) and making the reasonable assumption that the leptonic 
K\"ahler matrices $K^L, K^e$ and $K^\nu$ can be written as perturbations 
of the identity, $K^\phi=I+\Delta^\phi$, we can make a different type of 
expansion, in first orden of the perturbations $\Delta^\phi$. The analysis 
(and the results) resemble those of renormalization group 
evolution in cases with degeneracy \cite{RGEnu}.
For the charged lepton mass eigenvalues we obtain (no sum in $i$)
\be
m'_{e_i}\simeq m_{e_i}\left[1-{1\over 2}(V_{e_L}^\dagger\Delta^L 
V_{e_L})_{ii}
-{1\over 2}(V_{e_R}^\dagger\Delta^eV_{e_R})_{ii}\right]\ ,
\ee
while for the neutrino masses we get  (no sum in $i$)
\be
m'_{\nu_i}\simeq m_{\nu_i}\left[1-(V_{\nu_L}^T\Delta^L 
V_{\nu_L})_{ii}\right]\ .
\ee
Once again we find that the corrected masses are proportional to the naive 
ones.

For the MNS matrix we find
\be
V'_{MNS}\simeq V_{MNS} + X_e V_{MNS} - V_{MNS} X_\nu\ ,
\ee
where $X_e$ and $X_\nu$ are anti-hermitian with 
$(X_e)_{ii}=(X_\nu)_{ii}=0$ and
\bea
(X_e)_{ij} &= & {1\over m_{e_j}^2-m_{e_i}^2}
\left[m_{e_i} m_{e_j} (V_{e_R}^\dagger\Delta^eV_{e_R})_{ij}
+{1\over 2} ( m_{e_i}^2+m_{e_j}^2) (V_{e_L}^\dagger\Delta^L V_{e_L})_{ij}
\right]\ ,\nonumber\\
(X_\nu)_{ij} &= & 
{1\over 2} { m_{\nu_i}+m_{\nu_j}\over  m_{\nu_j}-m_{\nu_i}} 
(V_{\nu_L}^T\Delta^L V_{\nu_L})_{ij}\ .
\eea
It is straightforward to derive the second order corrections if they are 
needed. Here we simply notice from the first order result above that 
the corrections coming from the left-handed sector can indeed be large
in cases of (near)-degeneracy, as anticipated and do not attempt a 
more detailed analysis.

To finish this section, let us discuss briefly the case in which
the non-renormalizable dimension five operator in eq.~(\ref{Wlep})
comes from a see-saw mechanism: $\kappa=Y^{\nu} {\cal M}^{-1} {Y^{\nu}}^T$, 
where $Y^{\nu}$ is the neutrino Yukawa coupling and ${\cal M}$ 
is the right-handed Majorana mass matrix. It can be checked that
the canonical normalization of the right-handed neutrino superfields
does not affect the see-saw predictions for the low energy
neutrino mass matrix (or the non-renormalizable operator $\kappa$). 
Therefore  the analysis and results for this case 
are completely identical to the ones that we have just discussed
for the non-renormalizable operator.

\section{Conclusions}

In this paper we have analyzed the effect of the K\"ahler potential
on the Cabibbo-Kobayashi-Maskawa (CKM) matrix, taking into account
the flavour violation that it induces on the Yukawa matrices when
the kinetic terms are canonically normalized. We have derived
an exact formula [eq.~(\ref{newformCKM})] that relates the CKM matrix that 
one would naively
compute from the Yukawa couplings, to the
physical CKM matrix, computed redefining properly the quark superfields
to render the kinetic terms canonical. We have analyzed this
formula requiring only that the quark masses and mixing angles
are the observed ones, and we have proved
that the contributions to the CKM matrix from the K\"ahler
potential are subdominant. Such subdominance has been found previously
in some concrete models (see e.g. \cite{Leurer:1992wg,camu}), but it 
could always be 
attributed to the particular properties of the model under discussion. In 
contrast, the proof we have presented in this paper is 
model-independent.

We have also undertaken a similar analysis for the lepton sector,
to study the impact of the flavour mixing in the K\"ahler potential
on the Maki-Nakagawa-Sakata matrix. Our conclusions are different
depending on the neutrino mass spectrum. When neutrinos are hierarchical
we do not expect the K\"ahler potential to be the dominant contribution
to the mixing angles. At most they would give a contribution to the mixing
of the same order of the contribution from the superpotential. This
contribution could be important, though, to explain the deviations
from maximality in the solar and atmospheric mixings, for the case
in which the superpotential yields maximal mixing angles by itself.
On the other hand, when neutrinos are degenerate, important changes
in the mixing angles are expected, due to the sensitivity
of the mixing angles to any new effects, when the superpotential 
yields degenerate mass eigenstates.

\section*{Note added}

After the completion of this work, we learned about two
groups working along the same lines \cite{KPRVV}, \cite{JJ}. 
Their conclusions agree with ours, in the aspects
where our analyses overlap.

\section*{Acknowledgements}

We would like to thank Alberto Casas and Andrea Romanino for discussions, 
and Tim Jones and Oscar Vives for sharing their results with us before 
publication. J.R.E. thanks the CERN TH group for hospitality and partial 
financial support during the initial stages of this work.


\end{document}